\documentclass[12pt]{iopart}
\usepackage{graphics,graphicx,marvosym}
\usepackage{dcolumn,bm}
\usepackage{mathbbol}

\begin{document}

\title{Response of Two-dimensional  Kinetic Ising Model under Stochastic Field}
\author{Asim Ghosh$^1$ and Bikas K. Chakrabarti$^{1}$}
\address{$^1$Theoretical Condensed Matter Physics Division, 
Saha Institute of Nuclear Physics, 1/AF Bidhannagar, Kolkata 700 064 India}
\eads{asim.ghosh@saha.ac.in \& bikask.chakrabarti@saha.ac.in}


\begin{abstract}
We study, using Monte Carlo dynamics, the time ($t$) dependent average magnetization per spin  $m(t)$ behavior of  2-D kinetic  Ising model   under  a binary ($\pm h_0$) stochastic field $h(t)$.  The time dependence of the stochastic field  is  such that its average over each successive time interval $\tau$  is assured to be zero (without any fluctuation). The  average magnetization $Q=(1/\tau)\int_{0}^{\tau} m(t) dt$ is considered as order parameter of the system. The phase diagram in ($h_0,\tau$) plane is obtained.  Fluctuations in order parameter and their scaling properties are studied  across the phase boundary. These studies indicate that the nature of the transition is Ising like (static Ising universality class) for field amplitudes $h_0$ below some threshold value $h_0^c(\tau)$  (dependent on $\tau$ values; $h_0^c\rightarrow0$ as $\tau\rightarrow\infty$ across the phase boundary) . Beyond these $h_0^c (\tau)$, the transition is no longer continuous.  
\end{abstract}

\maketitle


\section{Introduction}
\noindent The study of  the kinetic Ising system under time varying magnetic field has  already been  an important research area of non-equilibrium statistical physics \cite{Binder1986,Gammaitoni1998,Chakrabarti1999}. In this context, many studies were made  where a periodic time varying magnetic field  was applied in the kinetic Ising system and it was observed that a  symmetry breaking dynamic transition (in the response magnetization) takes place depending upon the magnitudes  of frequency and amplitude of the applied field \cite{Lo1990,tome}. Extensive Monte Carlo simulations have been done  \cite{Acharyya1995,Chakrabarti1999,Rikvold1994,Rikvold1998} to estimate   the different  critical exponents for this dynamic transition. It  appears (see e.g., \cite{Chakrabarti1999}) that   the nature of dynamic transition  is Ising like (continuous) up to  a certain value of field amplitude and  frequency after which  the  transition becomes first order.  Of course, some of the later studies suggested \cite{Rikvold2000,Sides98,Korniss02} this to be a  finite size effect.   The same model was also studied by applying field pulse \cite{Misra1998,Chakrabarti,Chatterjee2003,Chatterjee2004}.

In this paper we will discuss the behavior of  kinetic Ising model under stochastic field  (random field in time; $h(t)$). The   kinetic Ising model under random field has also been addressed  earlier by taking different types of distributions of external magnetic field \cite{Chakrabarti1999, Acharyya1998, Hausmann1997}, though its transition behavior  has not been analyzed systematically.  Here we will investigate the response of  kinetic Ising model under binary stochastic field $\pm h_0$. We  study  the model by taking the stochastic field such that within every  successive time interval $\tau$,  the total field applied on the system is zero (without any fluctuation; $\int_{n\tau}^{(n+1)\tau}h(t)dt=0$ for any integer value of $n$). We did  Monte Carlo studies for different values of $\tau$. We observe that a continuous dynamic phase transition takes place for small  values of field amplitude  $h_0$ and small $\tau$ values with Ising-like scaling behavior and exponent values. However, for higher values of $\tau$ or   field amplitude $h_0$, the nature of transition does not remain continuous.

Specifically, we have studied the response magnetization $m(t)$ of the system for  stochastic fields $h(t)$ and define an order parameter $Q=(1/\tau)\int_{0}^{\tau} m(t) dt$ (where $m(t)$ is the average magnetization per spin). We study the fluctuation behavior and scaling properties for different ranges of $\tau$ values. In particular,  here we study the variation of the Binder cumulant 
$U_L$ ($=1-\frac{<Q^4>_L}{3<Q^2>^2_L}$, where $<\dots>$ denotes the thermal average in the steady state) and the fluctuation or susceptibility $\chi_L$ (=$(L^2/k_{B}T)[<Q^2>_L - <Q>^2_L]$ ; $T$ denotes  the temperature  and $k_{B}$ is the Boltzmann factor) at different values of system size $L$ and different values of $h_0$ and $\tau$.

We find that up to a threshold value of $h_0$ ($h_0^c(\tau)$, dependent on  $\tau$) the crossing points of $U_{L}$ for different $L$ values match with the extrapolated peak position in $\chi$ and the crossing point value $U^*$ of the Binder cumulant compares well with that of the pure static Ising value, indicating a continuous transition with identical universality class. Scaling behavior of $\chi$ also suggests that. Beyond the $h_0^c(\tau)$ value, however, a clear cross-over takes place. The phase diagram, giving  $h_0^c(\tau)$, is obtained ($h_0^c(\tau)\rightarrow0$ as $\tau\rightarrow\infty$).  Beyond the crossover, the Binder cumulant scaling behavior suggests an immediate drop from its complete order value seems to  indicate a first order like transition. The same is suggested by measurement of  the  susceptibility peak  values  $\chi_{max}$ for different system sizes ($L$), giving  $\chi_{max}\sim L^d$ ($d$ denotes dimension of the lattice). However   the estimated value of correlation length exponent $\nu$ seems to be quite high, seems to indicate this discontinuous transition from  the ordered phase, beyond $h_0^c(\tau)$, to be a `glass-like' dynamically frozen phase \cite{Hausmann1997,Rieger96}.


\section{The Model and Monte Carlo simulation}
\noindent We consider  two dimensional ($L\times L$ on a square lattice) kinetic Ising model with periodic boundary condition. The Hamiltonian of the system can be written as 
\begin{eqnarray}
 H=-\displaystyle\sum_{(ij)} J_{ij}s_i s_j -h(t)\sum_{i} s_i
\end{eqnarray}
where $J_{ij}$ is the interaction strength between the $i$-th and $j$-th spins (here we take $J_{ij}=1$), $s_i=\pm 1$ for any $i$-th spin, $(ij)$ indicates the nearest-neighbor pairs and $h(t)$ is  the field applied on the system. We consider  time dependent stochastic field $h(t)$ which varies stochastically over time and in our case it takes the values $\pm h_0$ with same probability.  In order to avoid case with a net average of $h(t)$ due to fluctuation, we choose the field amplitudes  in such a way that within a certain time interval $\tau$  total field applied in the system is zero: In each period $\tau$, a series of $\tau/2$ number of $h_0$ and $-h_0$ values are first chosen and then called randomly from the set. This ensure that $\int_{n\tau}^{(n+1)\tau}h(t)dt=0$ for all integer values of $n$.  The order parameter of the system can be defined as
\begin{eqnarray}
 Q=(1/\tau)\int_{t=n\tau}^{(n+1)\tau}m(t) dt~,
\label{eqn1}
\end{eqnarray}
averaged over $n$ ($=0,1,2 \dots$), where $m(t)=(1/L^2)\sum_i s_i(t)$. In our  Monte Carlo simulation, we selected any spin randomly and then it was flipped with rate $min[1,exp(-\Delta E/k_B T)]$ , where $\Delta E$ is the change in energy due to the spin flip, $T$  is the temperature of the system and $k_B$ is the Boltzmann factor. One Monte  Carlo (MC) step is defined as   $L^2$  spins  updated randomly.

In our simulation, we have taken  $L=32,64,128$ and $256$ (with periodic boundary conditions) and the initial conditions were  either all spins up or down. To check  the steady state, we took the averages  for some initial time (typically $10^5$ MC steps) and checked if the averages match at least for three to five  successive such time intervals. After that, the order parameter,  and  other parameters, discussed later, were   averaged  over more than $10^6$ time steps.          
\section{Results}
\begin{figure}[tbh]
\begin{center}
 \centering
 \includegraphics[width=0.5\linewidth]{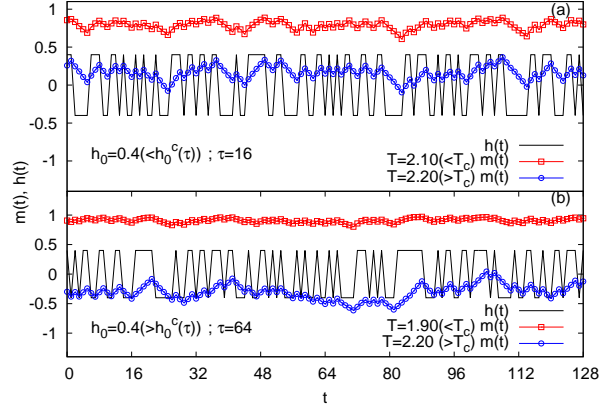}
\end{center}
   \caption{Time variations of average magnetization $m(t)$ in ordered phase and disordered phase  are shown here.  (a)$\tau=16$ and $h_0=0.4$ ($<h_0^c(\tau)$); $T=2.20$ corresponds to disordered phase while $T=2.10$ corresponds to ordered phase.  (b)$\tau=64$ and $h_0=0.4$ ($>h_0^c(\tau)$); $T=2.20$ corresponds to disordered phase while $T=1.90$ corresponds to  ordered phase.  The simulations were done for $L=128$.}
\label{mt-t.eps}
\end{figure}

\begin{figure}[tbh]
\begin{center}
 \centering
 \includegraphics[width=0.45\linewidth]{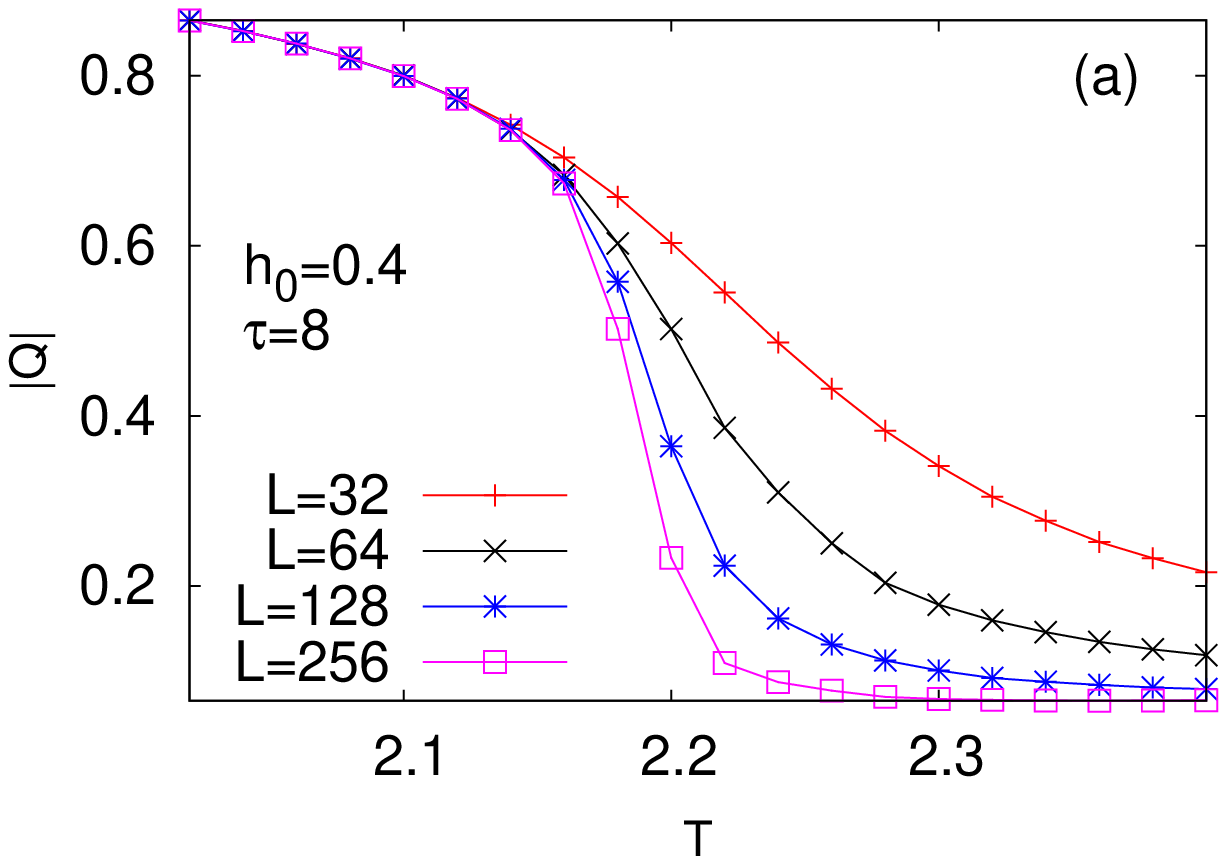}
\includegraphics[width=0.45\linewidth]{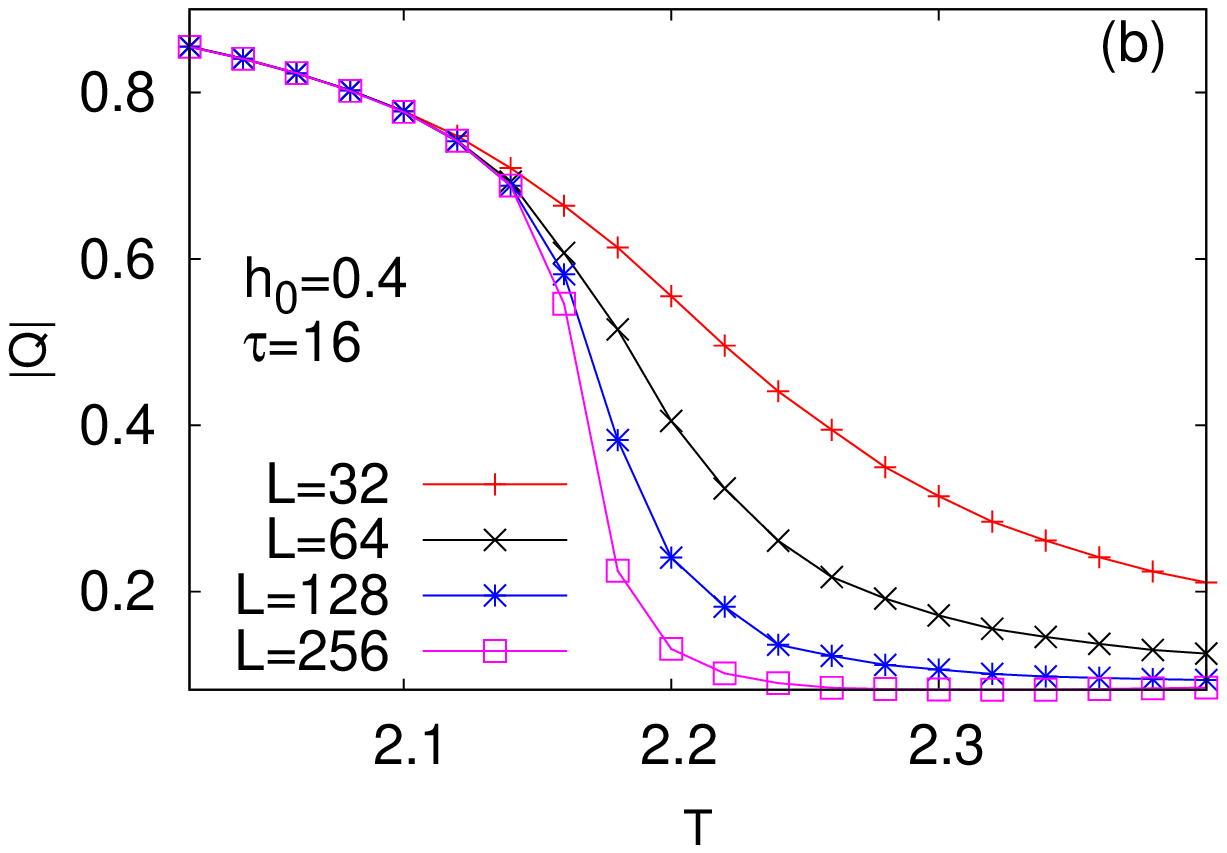}
\includegraphics[width=0.45\linewidth]{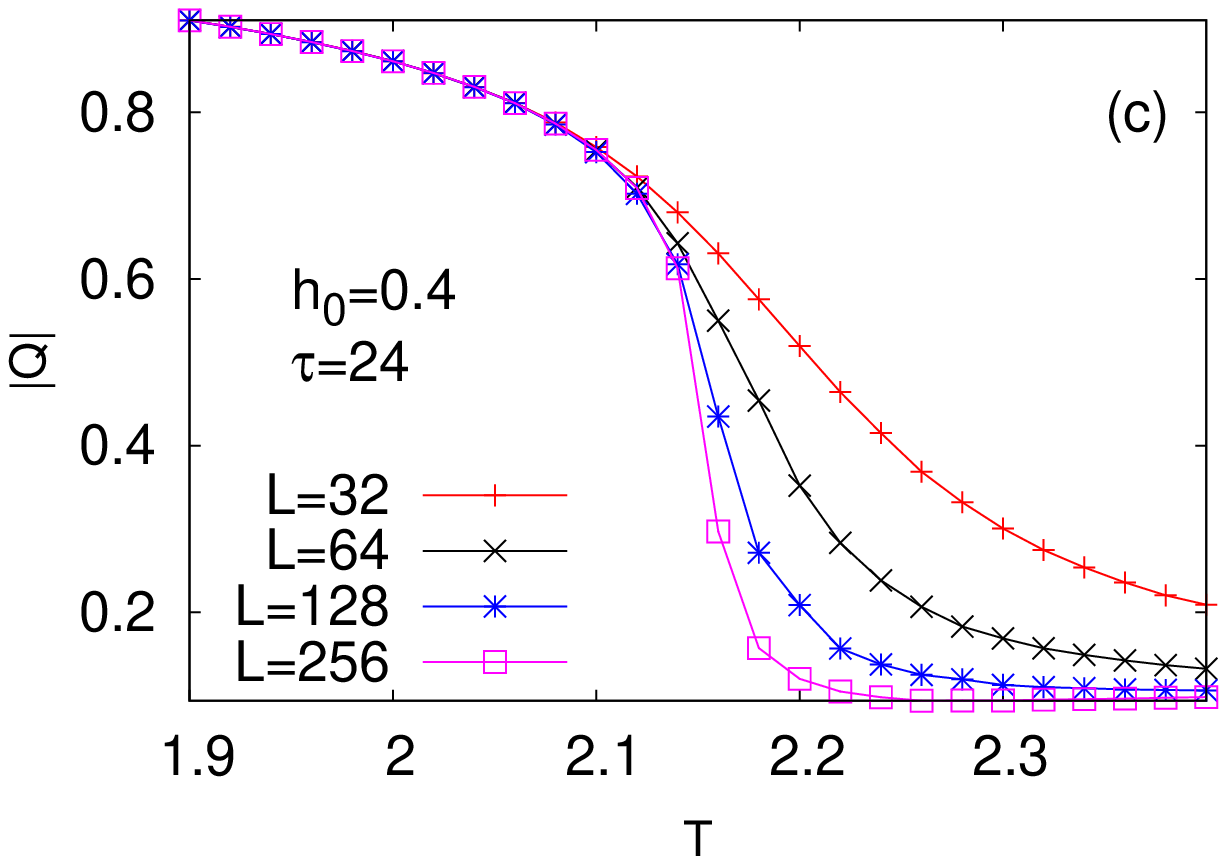}
\includegraphics[width=0.45\linewidth]{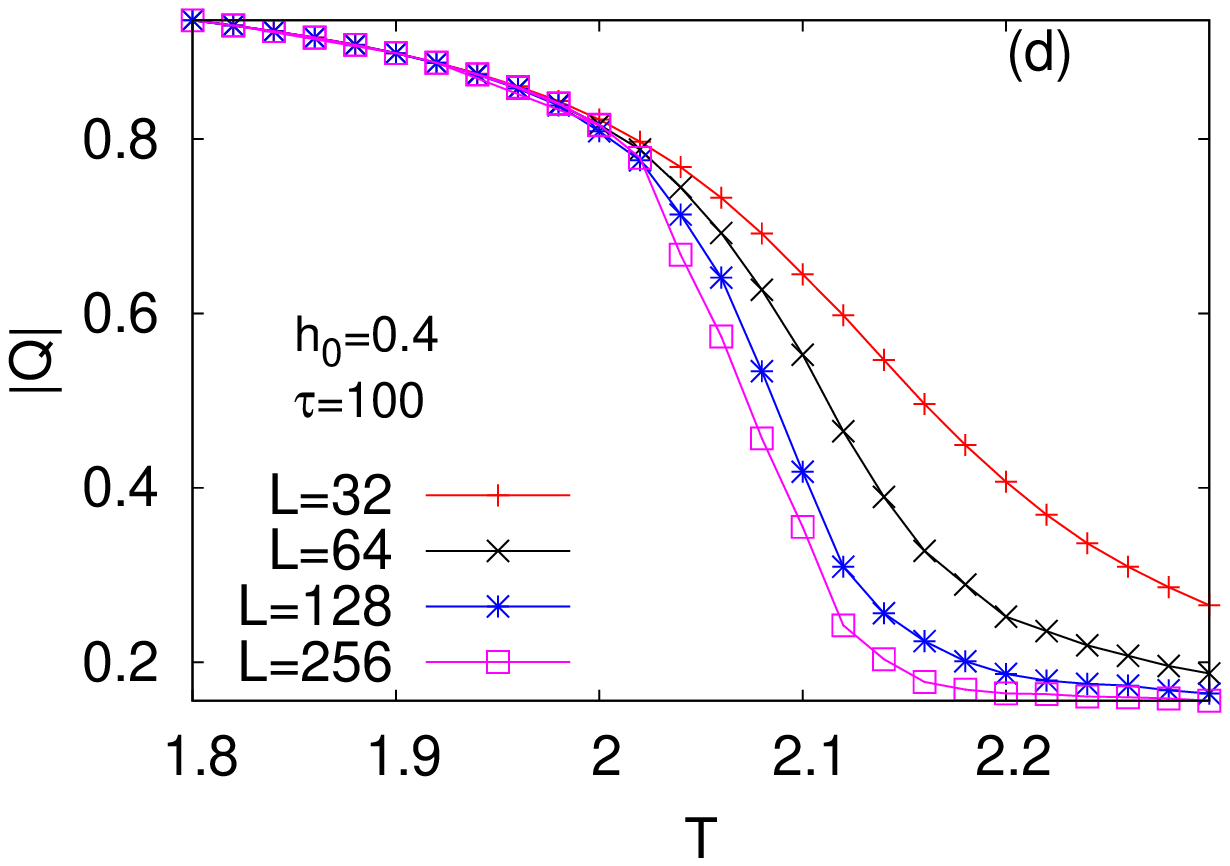}
\end{center}
   \caption{The order parameter for different system sizes are shown as function of temperature. (a) $h_0=0.4$ and $\tau=8$ (b) $h_0=0.4$ and $\tau=16$  (c)$h_0=0.4$ and $\tau=24$  (d)   $h_0=0.4$ and $\tau=100$.}
\label{order.eps}
\end{figure}

\noindent In Fig. \ref{mt-t.eps} we show the typical time variations of magnetization $m(t)$ for different values of field amplitude $h_0$ and time interval $\tau$. Transition from ordered state (with $Q\neq0$) to disordered state (with $Q=0$) as temperature changes is clearly seen.  We first study the temperature variation of the order parameter $Q$  for fixed value of  $h_0$ ($=0.4$) but for different $\tau$ values. In  Fig. \ref{order.eps}(a),  the order parameter variations with temperature are plotted  for different  $L$ values  for $\tau=8$.  The same for $\tau=16$,  $\tau=24$ and $\tau=100$  are shown in  Fig. \ref{order.eps}(b), \ref{order.eps}(c) and \ref{order.eps}(d)  respectively. 

\subsection{Binder cumulant behaviour}
\begin{figure}[tbh]
\begin{center}
 \centering
 \includegraphics[width=0.45\linewidth]{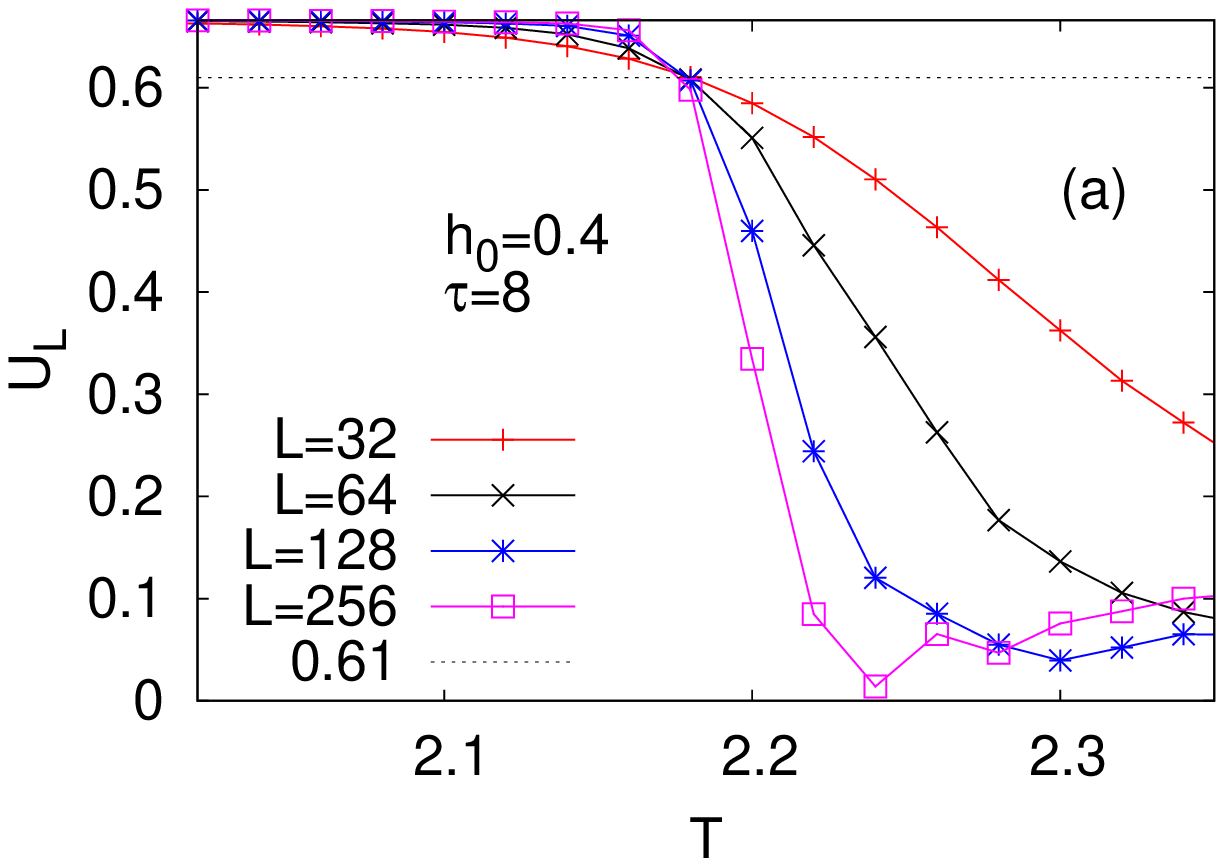}
\includegraphics[width=0.45\linewidth]{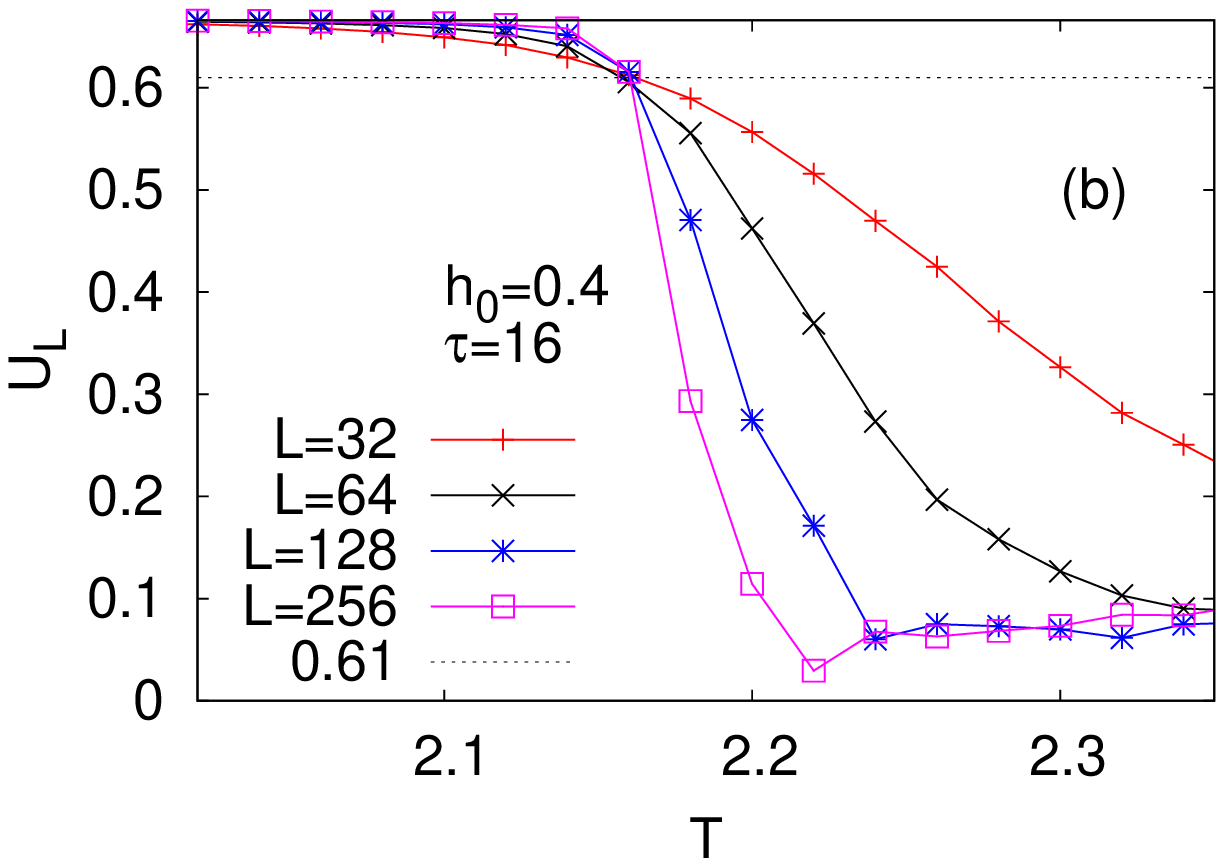}
\includegraphics[width=0.45\linewidth]{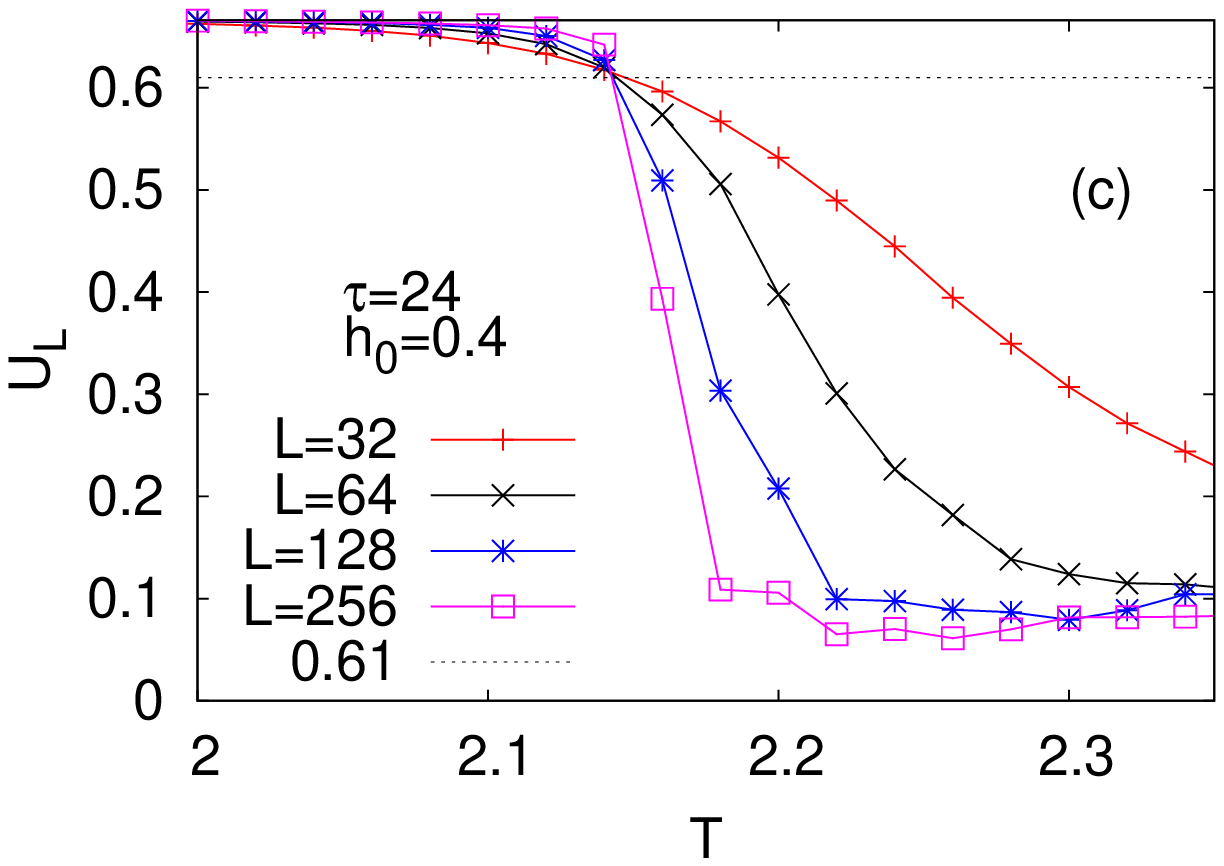}
\includegraphics[width=0.45\linewidth]{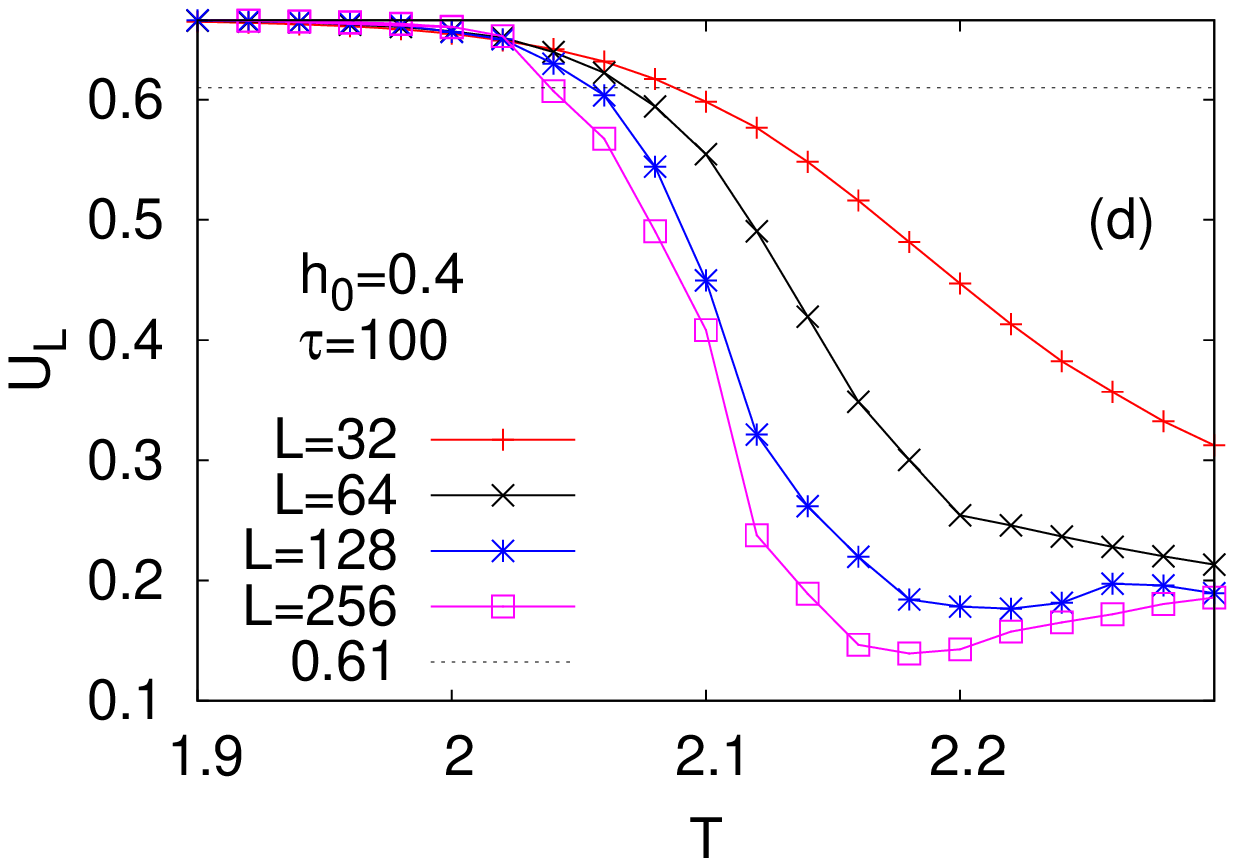}
\end{center}
   \caption{ (a) Temperature variation of the  Binder cumulant  for $h_0=0.4$ with different $\tau$ values are shown. (In Fig. (a) $\tau=8$  (b) $\tau=16$ (c) $\tau=24$   (d)  $\tau=100$.)  From the plots, we observe that $U^{*}$ is very much dependent on $\tau$ values: At higher $\tau$ values the   Binder cumulant  crossing  point $U^{*}$ deflects from value $0.61$ and eventually it assumes a value $0.66$.}
\label{binder.eps}
\end{figure}
\noindent To investigate  the transition,  we study  the Binder cumulant \cite{Binder1986} from the fluctuation of $Q$ for different $L$ values at fixed $h_0$ and $\tau$ values:
\begin{eqnarray}
 U_L=1-\frac{<Q^4>_L}{3<Q^2>^2_L}.
\end{eqnarray}
The temperature variations of the Binder cumulant for different sizes show a crossing pint $U^*(T_c)$ independent of system sizes at the critical point $T_c(h_0, \tau)$ (see e.g., \cite{Binder1986,Binder1988}).   In the Fig. \ref{binder.eps}(a) we plot $U_L$ for $L=32,64,128$ and $256$ with $\tau=8$. From the Fig. \ref{binder.eps}(a) we see that  there exists a  crossing point (for different $L$ values) with the value of  the cumulant $U^*\simeq0.61$ at a critical  temperature ($T_c\simeq2.18$ for $h_0=0.4$ and  $\tau=8$).  Fig. \ref{binder.eps}(b) and \ref{binder.eps}(c) correspond to  the same results for $\tau=16$ and $\tau=24$ respectively.  It is again observed that though the critical points change,  the Binder cumulant ($U^*$) does not change. Indeed this value of $U^*$ compares well with that for equilibrium Ising model, in 2-D.  In  Fig. \ref{binder.eps}(d), we have increased  the $\tau$ value further ($\tau=100$), and here we could not detect a  precise crossing point any where other than at the cumulant value for perfect order ($U_L=2/3$). The indication of transition after this perfectly  ordered phase suggests a cross over to first order transition in this large $\tau$ limit. 
\begin{figure}[]
\begin{center}
 \centering
\includegraphics[width=0.45\linewidth]{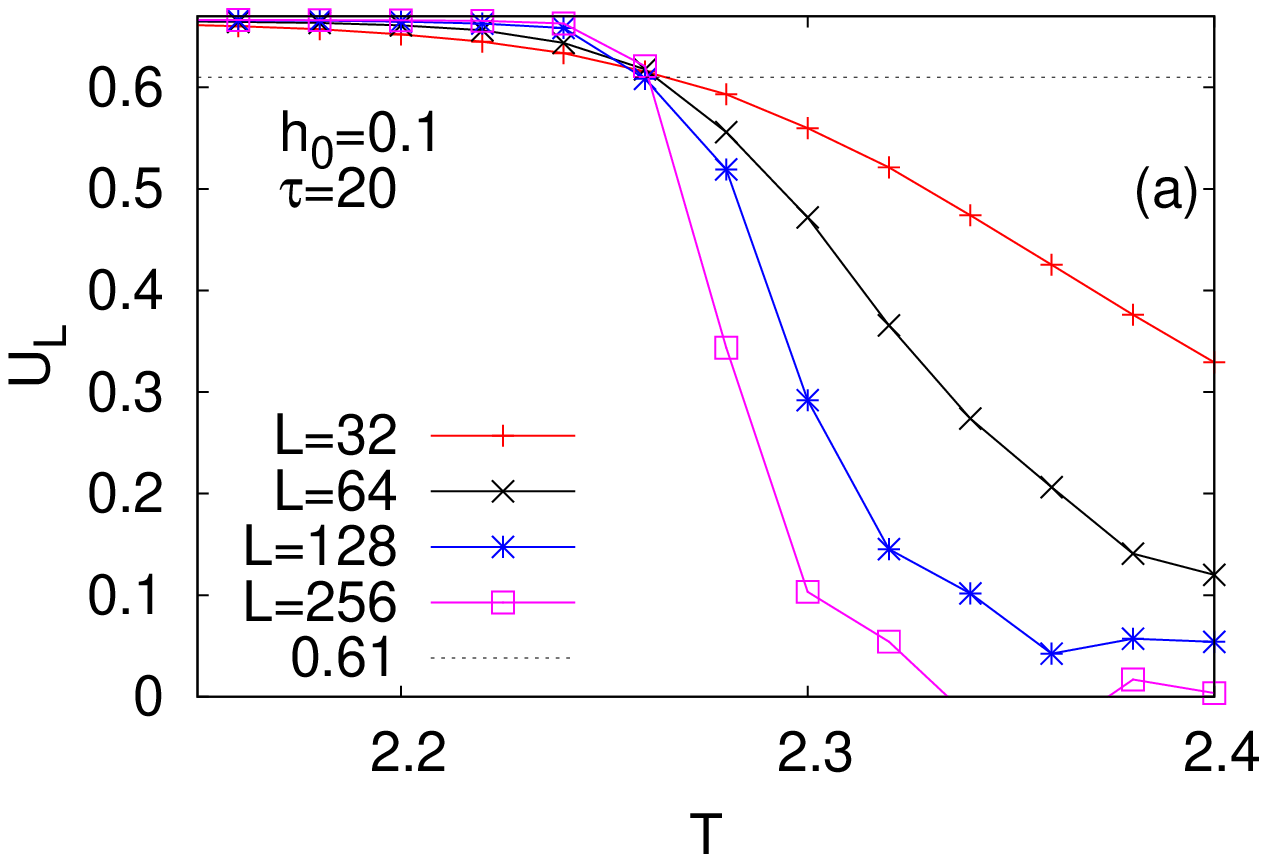}
\includegraphics[width=0.45\linewidth]{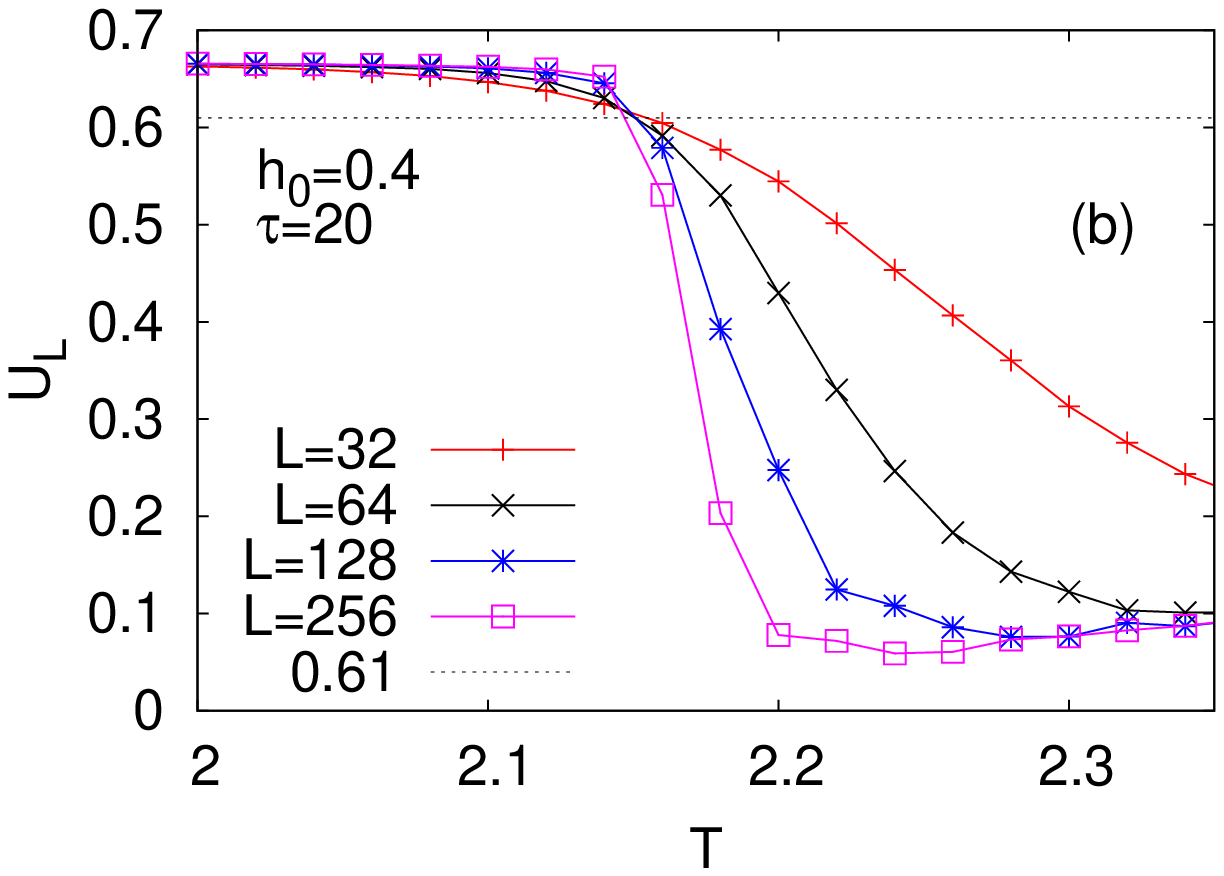}
\includegraphics[width=0.45\linewidth]{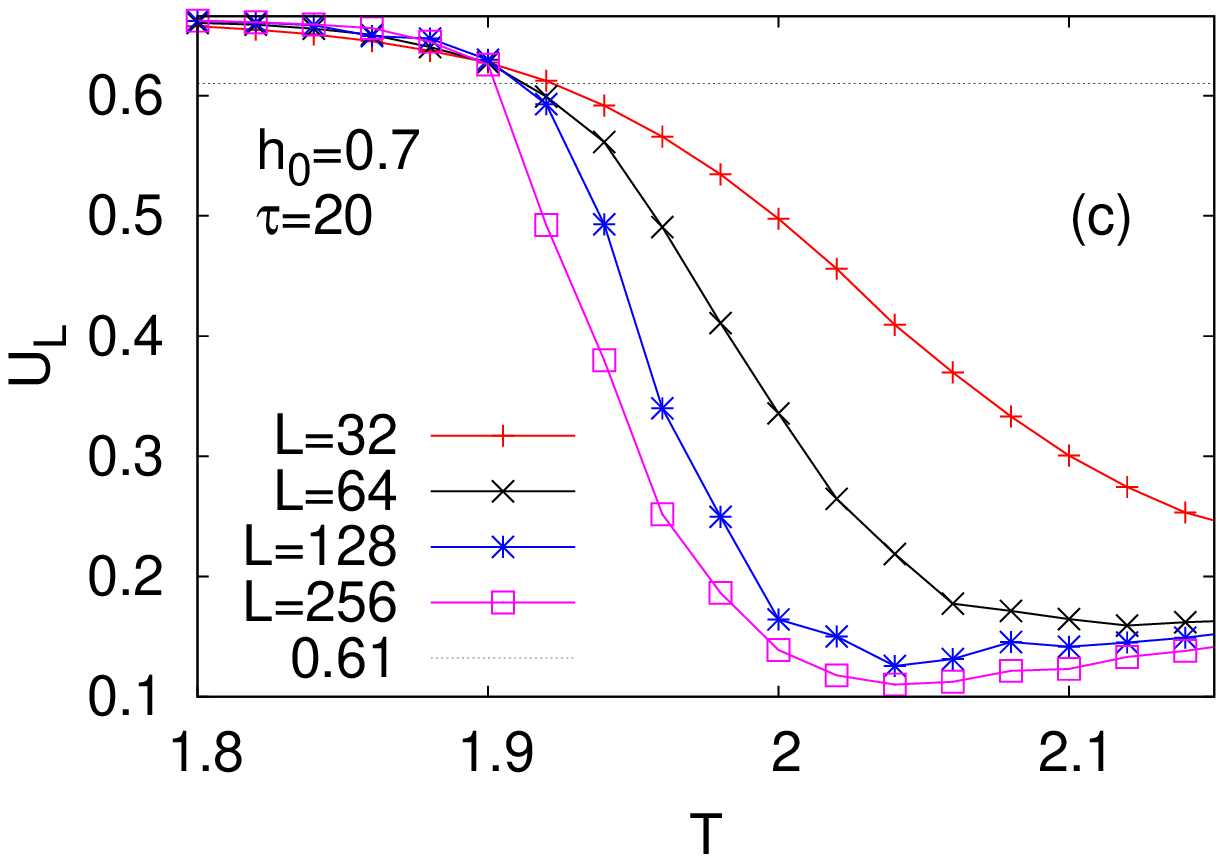}
\includegraphics[width=0.45\linewidth]{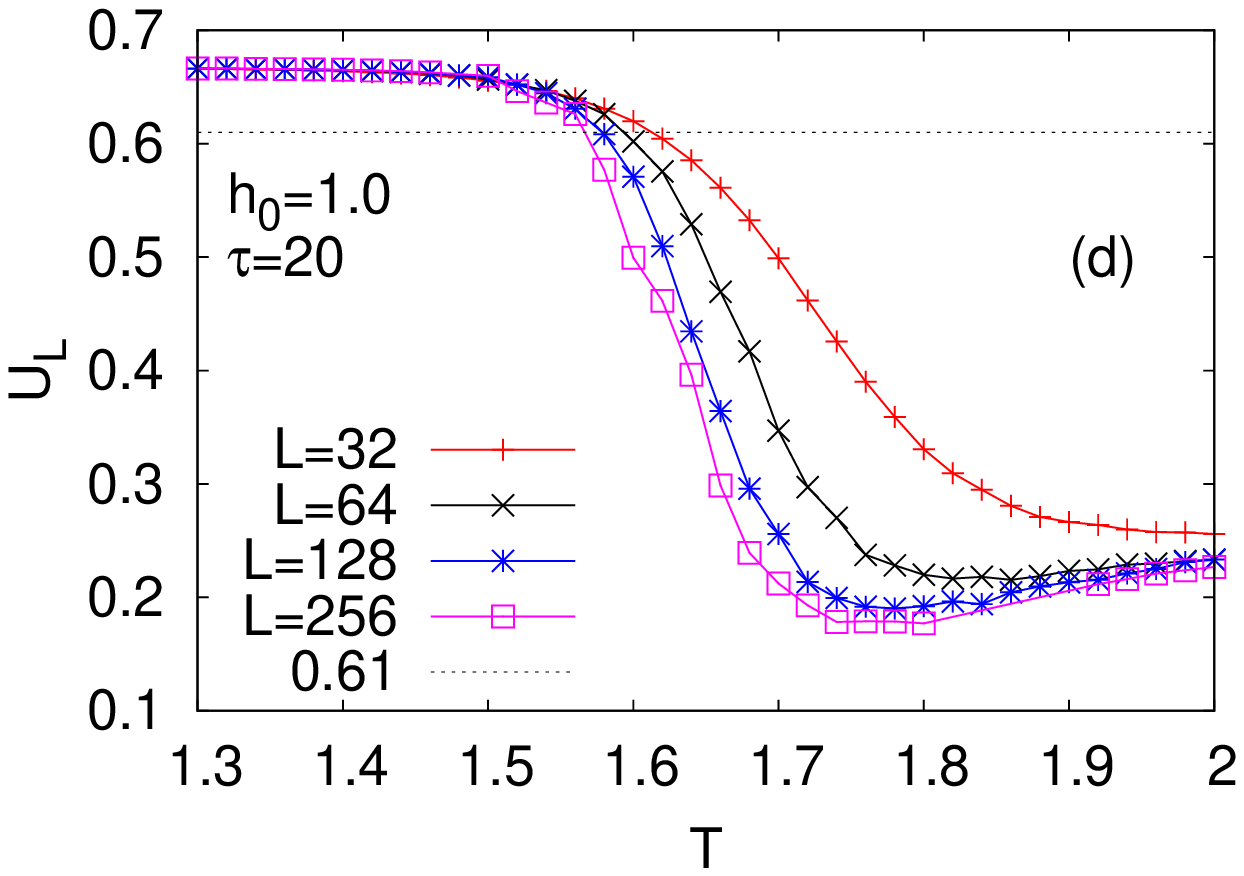}
\end{center}
   \caption{Binder cumulant for the same $\tau$ value ($=20$) but different $h_0$ values are shown in the figures (Fig. (a) $h_0=0.1$, (b) $h_0=0.4$, (c)  $h_0=0.7$ and  (d)  $h_0=1.0$). From the plots, we observe that $U^{*}$ is very much dependent on field values: At higher field values the   Binder cumulant  crossing  point $U^{*}$ deflects from value $0.61$ and eventually it assumes a value $0.66$, corresponding to completely order phase and seems to indicate a first ordered transition to the disorder phase (from complete order).}
\label{binder0.4-1.0.eps}
\end{figure}
Next, we  consider the fixed value of $\tau$ but different values of field amplitude $h_0$.   For example, we take four different field amplitude values  $h_0=0.1$, $0.4$, $0.7$ and $1.0$ with same $\tau$ ($=20$) value. Fig. \ref{binder0.4-1.0.eps} shows the  Binder cumulant values for system sizes $L=32, 64, 128$ and $256$. It is clear that the value of $U^*$ is around $0.61$ for smaller field amplitude (for $h_0\leq0.4$). But  in Fig. \ref{binder0.4-1.0.eps}(d) where the field amplitude ($h_0$) is taken as $1.0$, the value of $U^*$ is around $2/3$. Therefore, it is an indication that the static Ising universality like transition disappears for higher values of field amplitude $h_0$ as well as time period $\tau$.

\subsection{Phase Boundary} 
\begin{figure}[tbh]
\begin{center}
 \centering
 \includegraphics[width=0.6\linewidth]{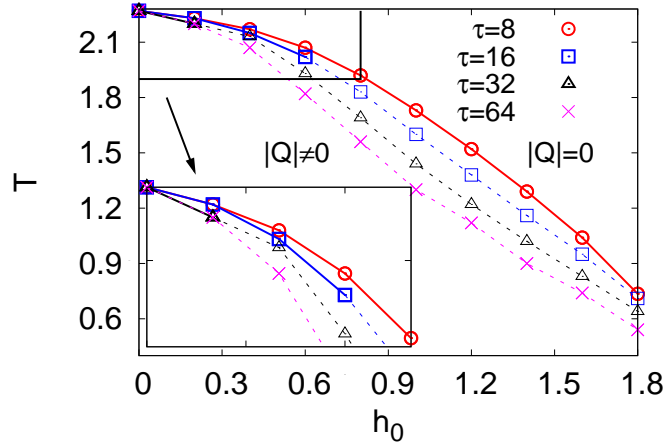}
\end{center}
   \caption{Dynamic phase boundaries are shown in the figure for $\tau=8$, $\tau=16$, $\tau=32$ and $\tau=64$. The
 nature of phase transition across the  boundary represented by solid line is static Ising universality class. For the phase boundary represented by dashed lines, nature of the  transition crosses over (from continuous  Ising like) to a discontinuous dynamically frozen spin glass like phase. The inset shows the crossover regions more clearly. The phase boundaries are obtained by measuring Binder cumulant crossing point for system sizes $L=32,64,128$ and $256$.}
\label{ran_phase.eps}
\end{figure}

\noindent
We have shown that there are certain ranges  of field amplitude  $h_0^c(\tau)$ for fixed value of $\tau$  below which the  nature of the transition is   static Ising universality class. But for $h_0>h_0^c(\tau)$ the nature of transition is completely different from static Ising universality. Here we give an effective phase boundary from order to disorder transition. Fig. \ref{ran_phase.eps} shows the phase diagram of ($h_0,T$) plane for different $\tau$ values. For any fixed value of $\tau$ ($=16$), there is a range of $h_0$ ($h_0^c(\tau)=0.6$) values below which the critical Binder cumulant ($U^*$)  is approximately $0.61$.    For $h_0>h_0^c$,  $U^*$ have much larger value (than $0.61$) and  it seems to  be $2/3$  corresponding to complete order. This part of the phase boundary is represented by dashed lines, across which a first order phase transition may occur. We find $h_0^c(\tau)\rightarrow0$ as $\tau\rightarrow\infty$ (practically $h_0^c\rightarrow0$ for $\tau>60$).
\subsection{Susceptibility and correlation length behaviors}
\begin{figure}[tbh]
\begin{center}
 \centering
\includegraphics[width=0.45\linewidth]{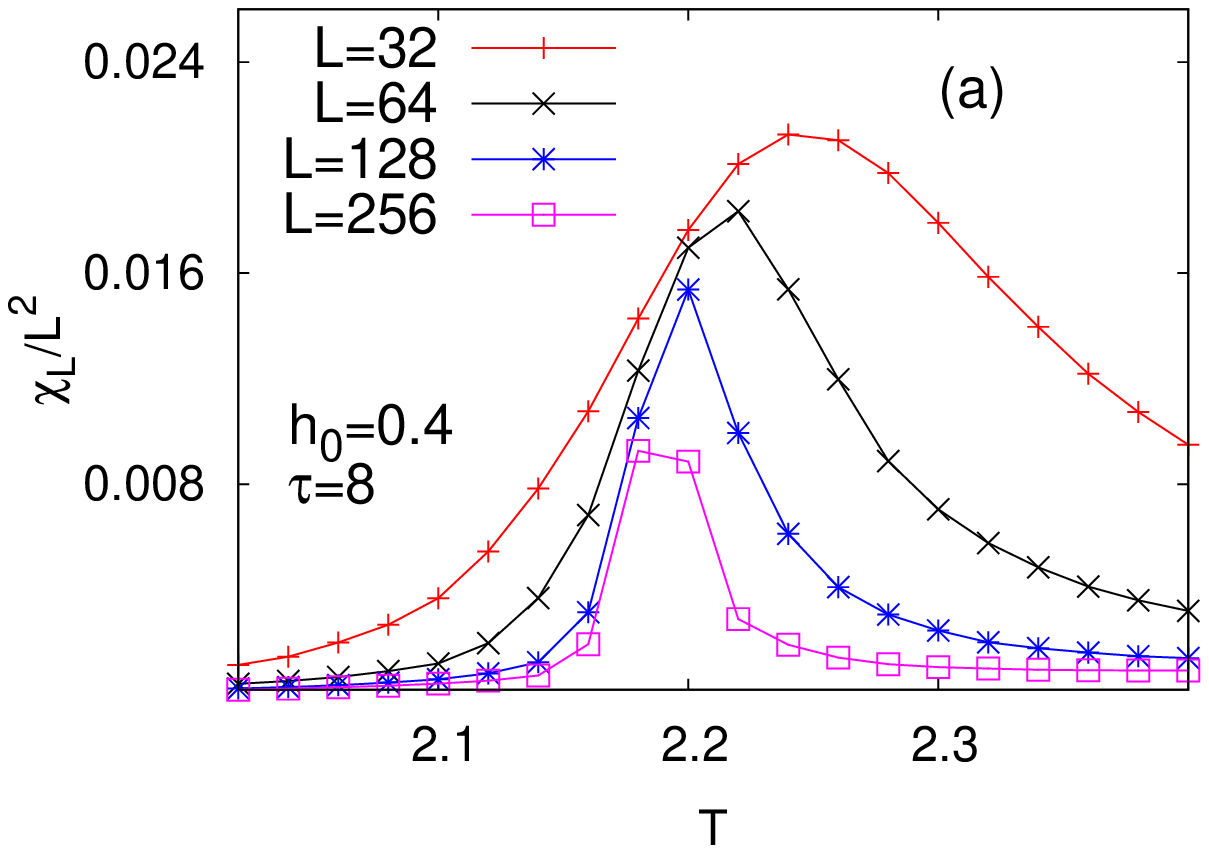}
\includegraphics[width=0.45\linewidth]{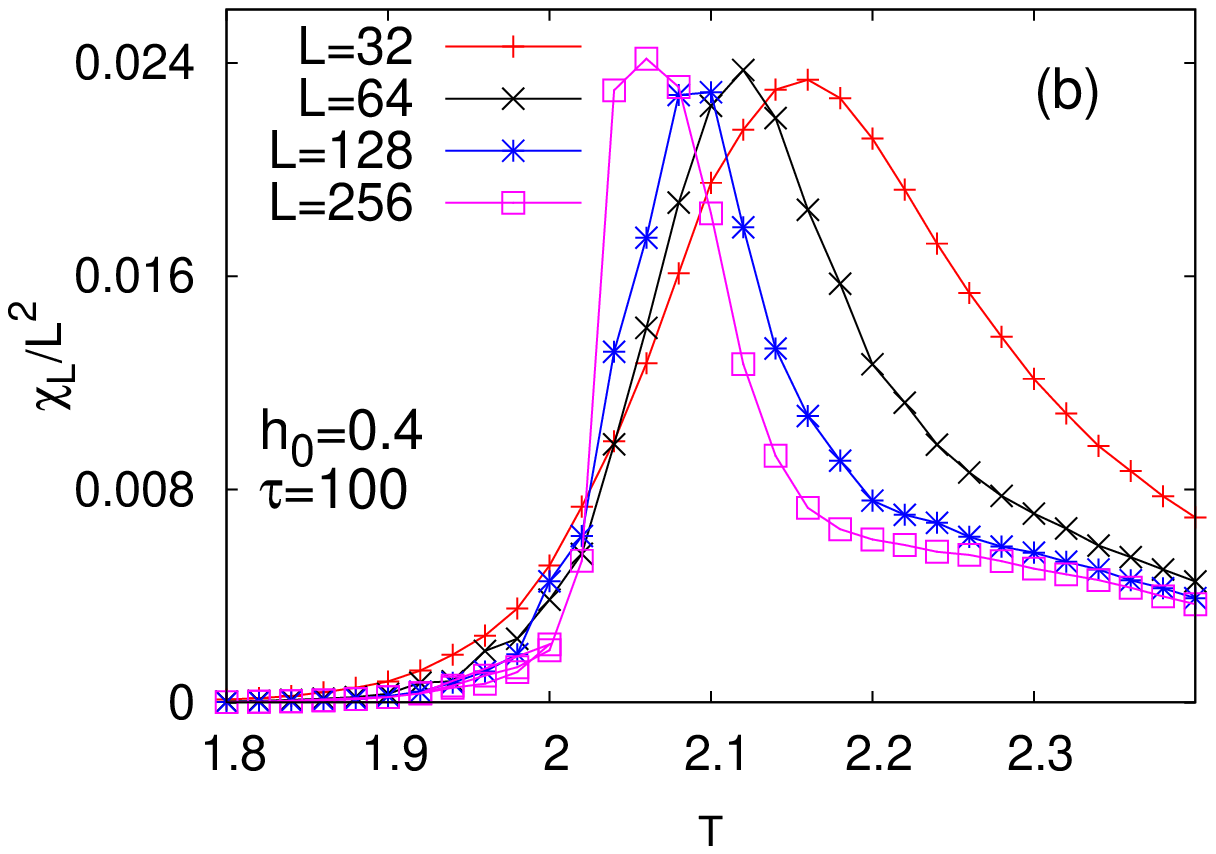}
\end{center}
   \caption{The temperature variation of susceptibility for different system sizes are  plotted for fixed field amplitude  $h_0=0.4$ but different $\tau$ values. (a) $\tau=8$ (b) $\tau=100$.}
\label{fluc.eps}
\end{figure}
\begin{figure}[tbh]
 \centering
 \includegraphics[width=0.48\linewidth]{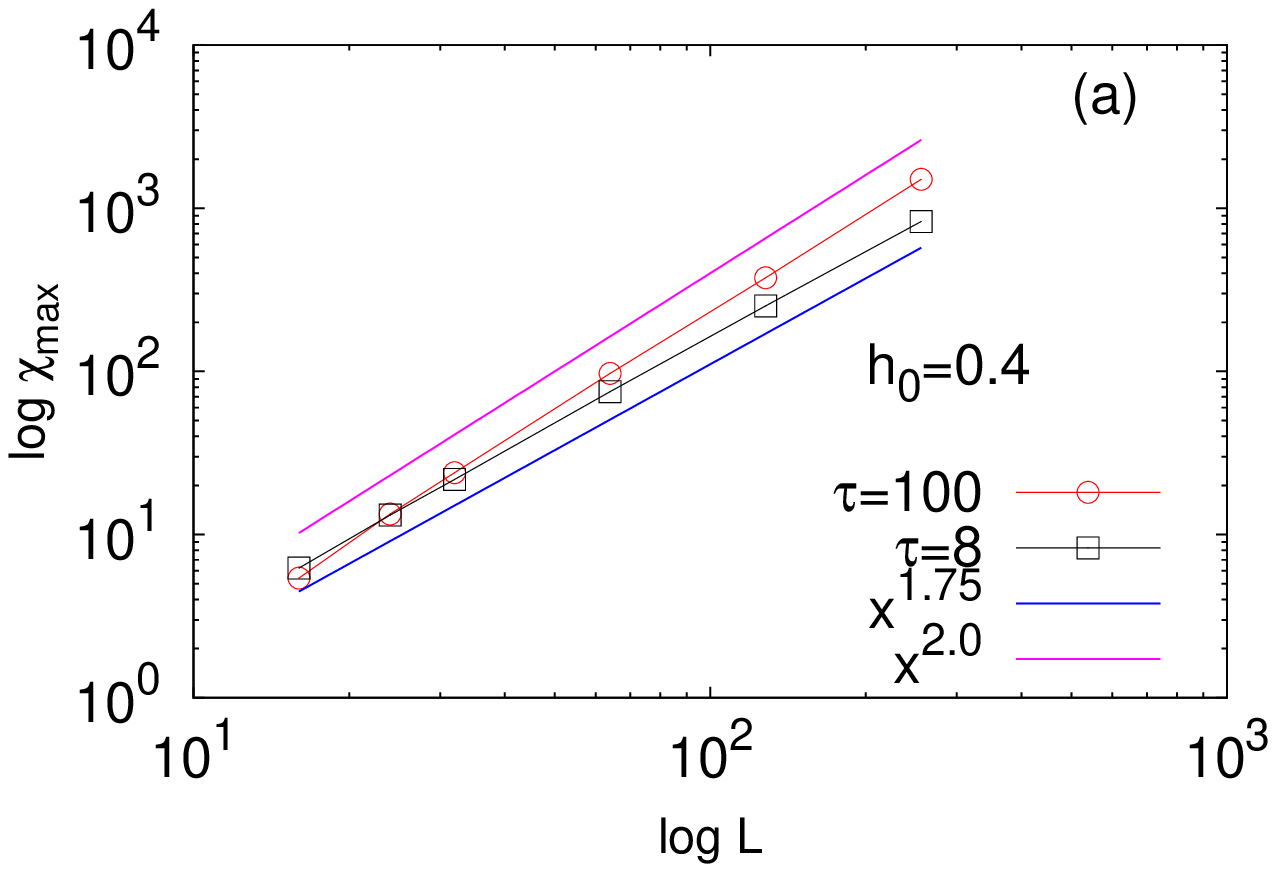}
\includegraphics[width=0.48\linewidth]{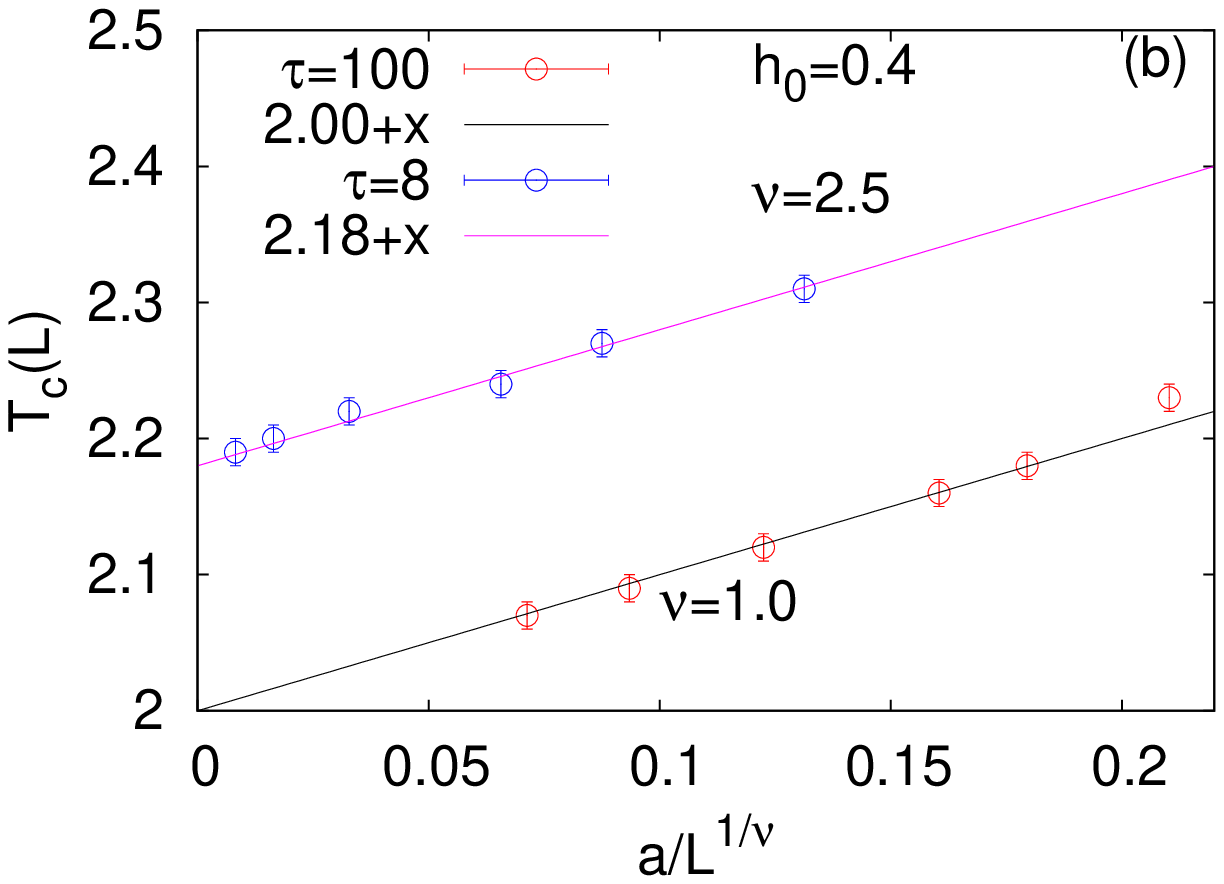}

   \caption{To estimate the critical exponents for lower range of $\tau$ (where $U^*$ is around $0.61$ ) and higher range of $\tau$ (where $U^*$ is around $2/3$) for fixed $h_0=0.4$,  we take $\tau=8$ and $\tau=100$. (a) The figure shows $\log-\log$ of  the system sizes versus the  maximum susceptibility. Assuming the scaling relation $\chi_{max}=L^{\gamma/\nu}$ we get $\gamma/\nu=1.75\pm0.05$ for $\tau=8$, and for $\tau=100$ the value is $2.0\pm0.1$.  (b)  Assuming the scaling relation defined in Eq. \ref{eqn2}, we calculate $T_c=2.18\pm0.01$ and $\nu=1.0\pm0.1$ for $\tau=8$. For $\tau=100$, it fits with this scaling relation by considering $T_c=2.00\pm0.01$ and $\nu=2.5\pm0.5$. For both figures, the simulations were done by taking $L=16,24,32,64,128$ and $256$.}
\label{cp-sm.eps}
\end{figure}
\noindent As mentioned earlier, we also investigated the behavior of fluctuations ($\chi$) in $Q$  for different system sizes. We define the  susceptibility as
\begin{eqnarray}
 \chi_{L}=(L^2/k_BT)(<Q^2>_L-<|Q|>^2_L).
\end{eqnarray}
and fit the data to the scaling form 
\begin{eqnarray}
 \chi_{L}=L^{\gamma/\nu}\chi^{0}(\epsilon L^{1/\nu}).
\label{eq:sus-scaling}
\end{eqnarray}
where the scaling function  $\chi^{0}$ is asymptotically defined  function and $\epsilon=(T-T_c)/T_c$. We estimated  $\chi_L$  for $L=32,64,128$ and $256$   for $\tau=8$ (Fig.  \ref{fluc.eps}(a)),  and  $\tau=100$ (Fig. \ref{fluc.eps}(b)) with  the  same field amplitude $h_0=0.4$. The peak point temperatures compare well with the estimates of $T_c(h_0,\tau)$ obtained from the Binder cumulant crossing point. 
To estimate  the critical exponents of the model for different values of $\tau$, we investigate the scaling  behavior of $\chi_L$.  As $\chi\sim\xi^{\gamma/\nu}$, where $\xi$ denotes the correlation length which is bounded by a maximum value of $L$ for finite systems, the  maximum value of  $\chi$ varies as $L^{\gamma/\nu}$. Fig. \ref{cp-sm.eps}(a) shows $\log-\log$ plot of the  susceptibility peaks as function of system sizes for $\tau=8$ and $\tau=100$.  It is  observed that  for $\tau=8$, the $\gamma/\nu$ fits with value $1.75\pm0.05$ which is very close to equilibrium Ising exponent value   $\gamma/\nu$ \cite{Binder1988}.  This result again  supports that the critical exponents  of the transition for smaller  values  of $\tau$ is close to Ising universality class. However,  for  larger values of $\tau\geq100$, the slope fits to  different  values (around $2.0$;  not comparing well  with  Ising universality class value). Again, this indicates perhaps a cross-over to first order transition for large $\tau$ values.  

To estimate  the correlation length exponent value of $\nu$ independently, we assume the scaling form as
\begin{eqnarray}
 T_c(L)=T_c-a L^{-1/\nu} 
\label{eqn2}
\end{eqnarray}
where $T_c$ is the critical point in thermodynamic limit and $T_c(L)$ is the critical point of system size $L$ and $a$ is any constant value. In the Fig. \ref{cp-sm.eps}(b) we plot $T_c(L)$ versus $a/L^{\nu}$ for $\tau=8$ and $\tau=100$. We observe that $\nu\simeq1$ (same as  equilibrium  2-D Ising exponent value) fits for $\tau=8$ and $h_0=0.4$ very well,  while  for $\tau=100$ and $h_0=0.4$, we get $\nu\simeq2.5$ as best fit value.  The estimated values of $\nu$ and $\gamma/\nu$   suggest both  that nature of phase transition is Ising like for portions of the phase boundary where $h_0<h_0^c(\tau)$ and first order (glass like \cite{Rieger96}) for $h_0>h_0^c(\tau)$.

\subsection{Scaling collapse for $h_0<h_0^c(\tau)$}
\begin{figure}[tbh]
 \centering
 \includegraphics[width=0.48\linewidth]{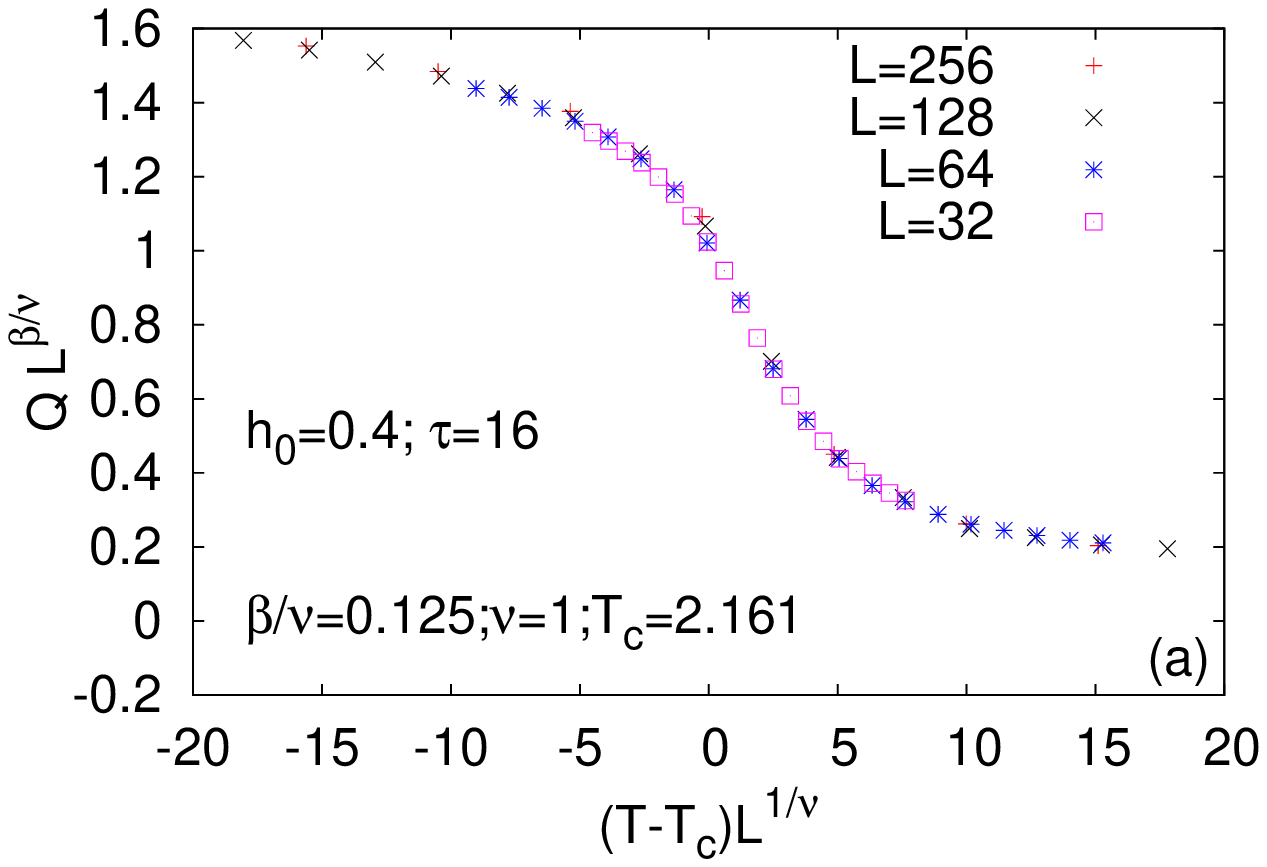}
\includegraphics[width=0.48\linewidth]{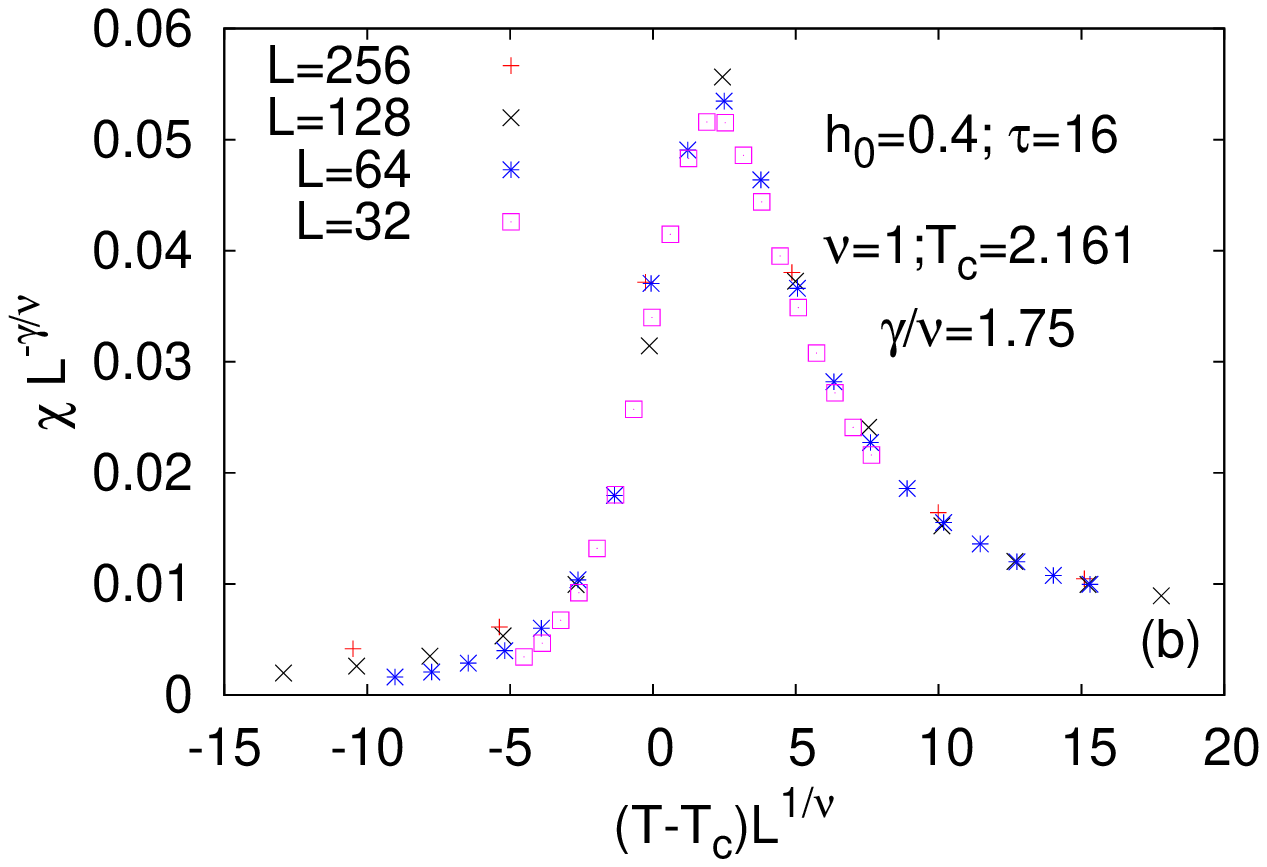}
\caption{(a) Data collapse for order parameter for different system sizes (b) Data collapse for susceptibility  for different system sizes. Both figures are simulated by taking $\tau=16$ and $h_0=0.4$ ($<h_0^c(\tau)$).}
\label{fig:coll}
\end{figure}

Our study here indicates that for field amplitude $h_0$ less than   a critical value of field amplitude $h_0^c(\tau)$, dependent on $\tau$, the transition belongs to  static Ising universality class.  To confirm,  we looked for  scaling collapse of data for $Q$ the order parameter  near critical point.  We assume that the order parameter  $Q$ and susceptibility $\chi$ scale  near critical point as 
\begin{equation}
 Q=L^{-\beta/\nu}Q^{0}(\epsilon L^{1/\nu}), ~~~ \chi=L^{\gamma/\nu}\chi^{0}(\epsilon L^{1/\nu}),
\end{equation}
where the scaling functions  $Q^{0}$ and $\chi^{0}$ are asymptotically defined  functions. We measured temperature variation of the average values of $Q$ for different system sizes for a given $\tau=16$ and $h_0= 0.4~(<h_0^c(\tau))$ and  looked for scaling fit. From data collapse, the estimated exponent values are  $\beta/\nu=0.125\pm0.005$ and $\nu=1.00\pm0.02$ (see Fig. \ref{fig:coll} (a)). These exponent values fit well  with the Ising universality class. We also found  data collapse for susceptibility and   our  estimated  exponent values are  $\gamma/\nu=1.75\pm0.05$ and $\nu=1.00\pm0.02$ (see Fig. \ref{fig:coll} (b)),  which are again  close to those for static Ising  universality class. 

For $h_0>h_0^c(\tau)$, no such good fit could be obtained (and that too for different exponent values if fitted)
and the Binder cumulant data indicated (see e.g., Fig. \ref{binder0.4-1.0.eps} (d)), a discontinuous drop of the cumulant value from about $0.66$ (complete order) to zero as temperature is increased beyond the  critical value.

\subsection{Absence of continuous transition for $h_0>h_0^c(\tau)$}
\begin{figure}[tbh]
\begin{center}
\centering
\includegraphics[width=0.45\linewidth]{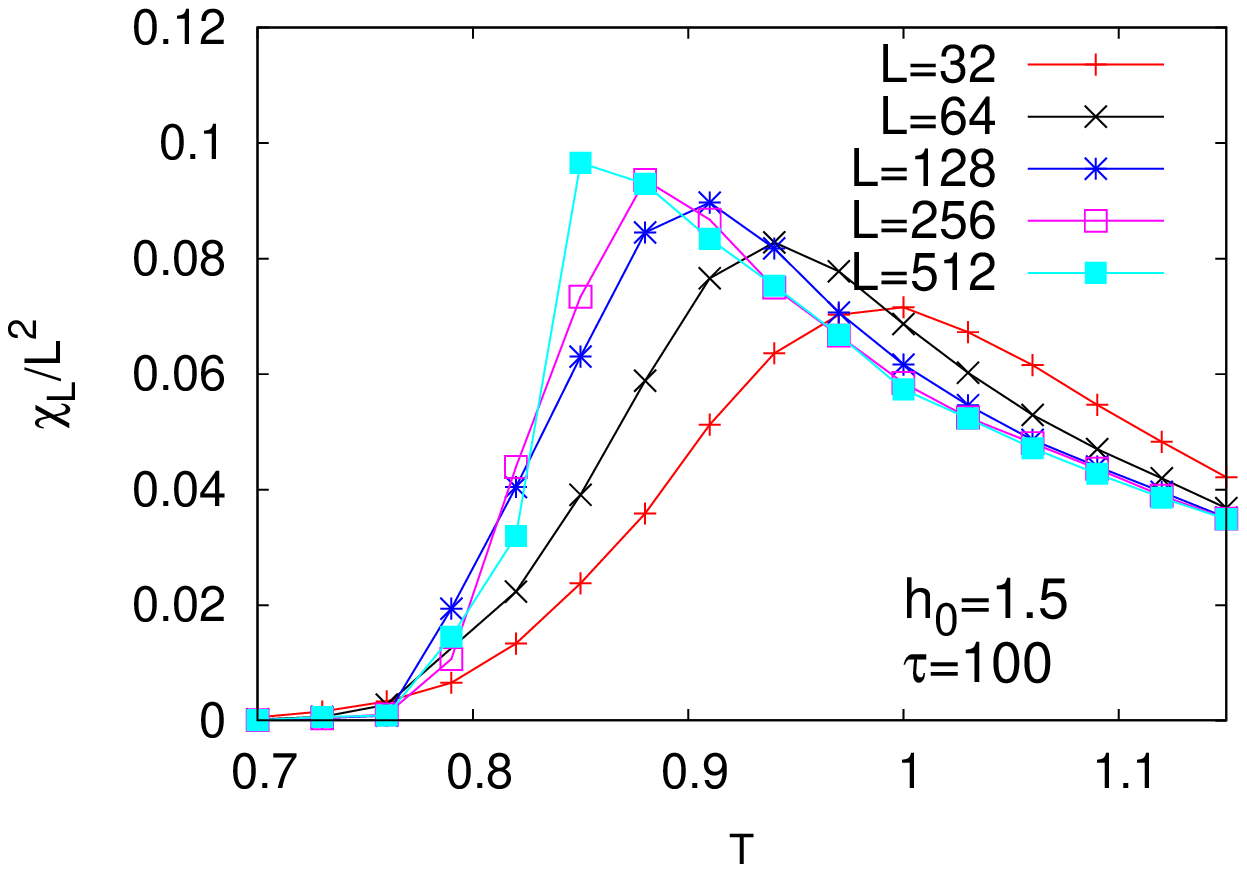}
\includegraphics[width=0.45\linewidth]{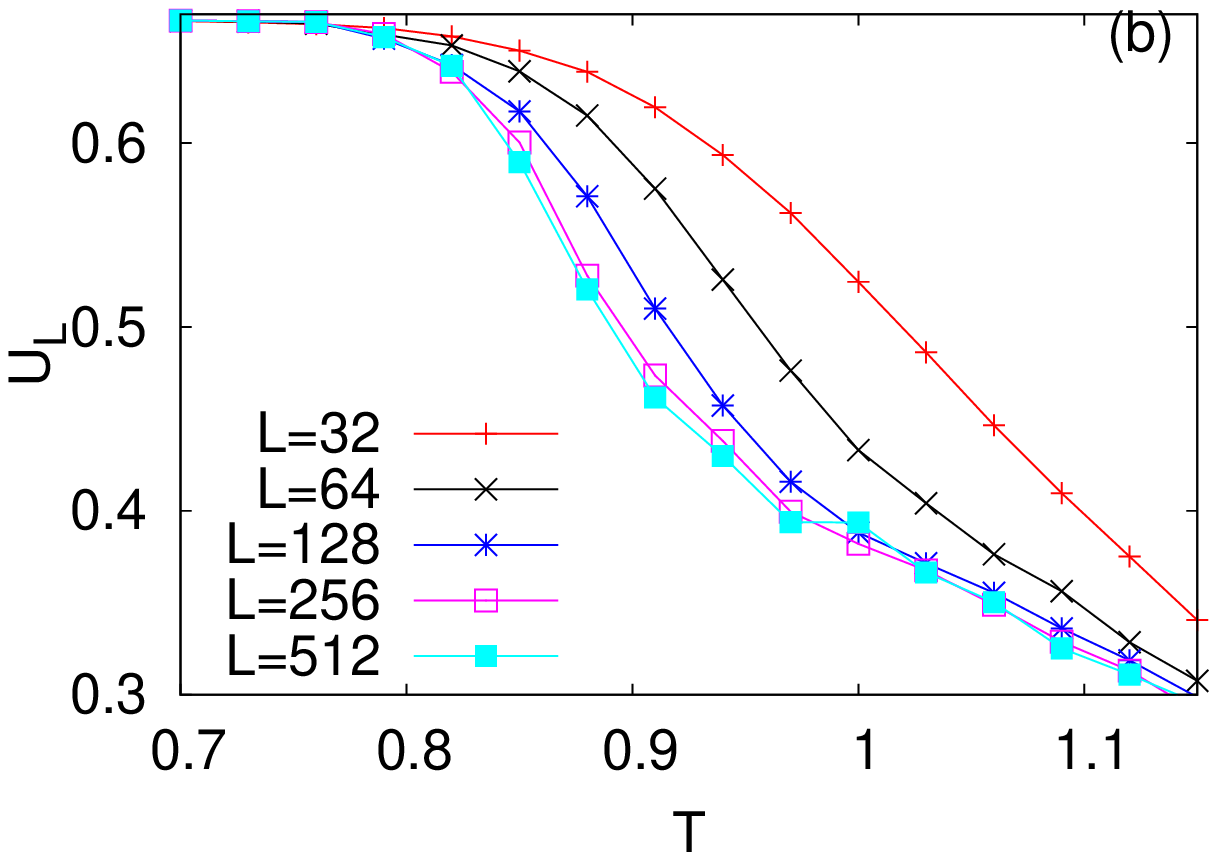}
\end{center}
   \caption{(a) The susceptibility  for different system sizes are plotted. From the figure it is clear  $\gamma/\nu\geq2$ (assuming $\chi_{max}\sim L^{\gamma/\nu}$).  (b) Here we plot Binder cumulant for different system sizes. Again from the figure it is clear that Binder cumulant  crossing point $U^* \simeq0.66$ (which is much different from static Ising value; may be  a signature of a 1st order transition).}
\label{new-fig}
\end{figure}

Here we have done the system size analysis for larger  sizes  where the systems  are claimed to be  showing 1st order or glass like transition ($h_0=1.5$ and $\tau=100$).  We plot susceptibility  for system different sizes ($L=32,64,128,256$ and $512$) as shown in Fig. \ref{new-fig}(a) and from the figure it is clear that  $\gamma/\nu>2$. We also plot the Binder cumulant correspondingly and from Fig. \ref{new-fig}(b)  a different behavior from  a 2nd order transition case is clearly seen.  

\section{Summary and Discussion}
\noindent Here  we have investigated, using Monte Carlo dynamics, a 2-D Ising spin system (on a square lattice with periodic boundary condition) under stochastically varying field (with binary values $\pm h_0$) and the spins are flipped according to  Glauber dynamics. Such systems were already considered earlier \cite{Chakrabarti1999,Acharyya1998,Hausmann1997} while the numerical study  \cite{Acharyya1998} was not very  conclusive,  the mean field study \cite{Hausmann1997} indicated several intriguing phases and transitions in  the model. In order to set a time scale for the stochastically varying  external field, we have time ordered the field in such a way that after every  time interval $\tau$ the total field applied in the system is zero i.e.,  $\int_{t=n\tau}^{(n+1)\tau} h(t)=0$ where $n=0,1,2,\dots$ and corresponding order parameter is defined as $Q=(1/\tau)\int_{t=n\tau}^{(n+1)\tau} m(t) dt$. To locate the critical point of the system precisely, we have measured the Binder cumulant for different system sizes. We have obtained  the phase diagram  in $h_0,T$ plane for different $\tau$ values and found out the cross over point between static Ising transition and non-static Ising transition. For truly stochastic field ($\tau\rightarrow\infty$), we find $h_0^c$ goes to zero. It has been observed that for  $h_0$ values less than $h_0^c(\tau)$,  given by  the phase boundary (in Fig. \ref{ran_phase.eps}),  the value of the Binder cumulant ($U^*$) is approximately $0.61$ (which fits well with corresponding value for static Ising universality class). For $h_0$ values greater than $h_0^c(\tau)$,  the Binder cumulant for different system sizes cross each other at a larger value, which is close to the value for the completely ordered state, indicating a discontinuous transition after that. We have also made a scaling analysis of the fluctuations of the order parameter $Q$ for this model.  We have seen that for $h_0$ values less than $h_0^c(\tau)$,  the maximum susceptibility ($\chi_{max}$) scales with $L^{\gamma/\nu}$ with scaling exponent $\gamma/\nu=1.75\pm0.05$ for different system sizes $L$ indicating Ising universality behavior. But for $h_0>h_0^c(\tau)$ the scaling exponent  $\gamma/\nu$  fits well to a value close to  $2.0=d$, seems to  indicate again a first order transition \cite{nuno2009,fisher82}. We also find that the correlation length exponent for the discontinuous transition has got unusually  high value ($\nu\simeq2.5$) seems to indicate a `spin-glass' like \cite{Rieger96} `frozen' dynamical \cite{Hausmann1997} phase.  

As suggested earlier \cite{Sides98,Korniss02} in the periodic field case \cite{Chakrabarti1999,Lo1990,tome, Acharyya1995},  this first order transition behavior (for $h_0>h_0^c(\tau)$, $h_0^c\rightarrow0$ as $\tau\rightarrow\infty$) may be a finite size effect. Our investigation,  so far, does not indicate of course any such finite size effect. For $\tau\rightarrow\infty$, however, the successive fluctuations will destroy any order and the system  is always in disordered phase.

\ack We thank A. Chatterjee for useful comments and suggestions. We would also  like to thank  S. Biswas  for discussions and critical reading of the manuscript.

\vspace{0.5cm}


\end{document}